\documentclass
[
aps,
pra,
twocolumn,
superscriptaddress,
titlepage,
secnumarabic,
nofootinbib,
nobibnotes,
raggedbottom,
showpacs
]
{revtex4-1}
\usepackage
{natbib}

\usepackage{hyperref}
\usepackage{graphicx}
\usepackage{amsmath}
\usepackage{amssymb}
\usepackage{braket}
\usepackage[retainorgcmds]{IEEEtrantools}
\usepackage{bbold}
\usepackage{multirow}
\usepackage{subfig}
\usepackage{color}

\definecolor{oxblue}{RGB}{0., 33, 71}
\definecolor{camblue}{RGB}{163,193,173}
\definecolor{BRGreen}{RGB}{1, 66, 37}
\definecolor{ngreen}{rgb}{0.2,0.6,0.2}

\newcommand{\ud}{\,\mathrm{d}}
	
\newcommand{\eulnum}{\mathrm{e}}
\newcommand{\im}{\mathrm{i}}

\begin{document}
\title{In Situ Characterisation of an Optically Thick Atom-Filled Cavity}

\author{J. H. D. Munns}
\email{joseph.munns@physics.ox.ac.uk}
\affiliation{Clarendon Laboratory, University of Oxford, OX1 3PU, United Kingdom}
\affiliation{Blackett Laboratory, Imperial College London, SW7 2AZ, United Kingdom}

\author{C. Qiu}
\affiliation{Clarendon Laboratory, University of Oxford, OX1 3PU, United Kingdom}
\affiliation{Department of Physics, Quantum Institute for Light and Atoms, State Key Laboratory of Precision Spectroscopy\\
East China Normal University, Shanghai 200062, Peoples Republic of China}

\author{P. M. Ledingham}
\affiliation{Clarendon Laboratory, University of Oxford, OX1 3PU, United Kingdom}

\author{I. A. Walmsley}
\affiliation{Clarendon Laboratory, University of Oxford, OX1 3PU, United Kingdom}

\author{J. Nunn}
\affiliation{Clarendon Laboratory, University of Oxford, OX1 3PU, United Kingdom}

\author{ D. J. Saunders}
\affiliation{Clarendon Laboratory, University of Oxford, OX1 3PU, United Kingdom}

\date{\today}
\begin{abstract}
A means for precise experimental characterization of the dielectric susceptibility of an atomic gas inside and optical cavity is important for design and operation of quantum light matter interfaces, particularly in the context of quantum information processing.  Here we present a numerically optimised theoretical model to predict the spectral response of an atom-filled cavity, accounting for both homogeneous and inhomogeneous broadening at high optical densities. We investigate the regime where the two broadening mechanisms are of similar magnitude, which makes the use of common approximations invalid.  Our model agrees with an experimental implementation with warm caesium vapor in a ring cavity.  From the cavity response, we are able to extract important experimental parameters, for instance the ground state populations, total number density and the magnitudes of both homogeneous and inhomogeneous broadening.
\end{abstract}

\maketitle

Reversible coupling between light and matter is an important element of optically based or mediated quantum information processing networks \citep{Ladd2010}.  Strong interactions are possible in a variety of media, utilising the field enhancements provided by cavities or the collective enhancement provided by interaction with large atomic ensembles, or both.  Examples include isolated atoms in high-Q cavities \citep{Boozer2007}, or atomic ensembles either in solid state \citep{Saglamyurek2011}, cold atomic gases \citep{Peyronel2012} or warm vapors \citep{Hosseini2011_nphys}. A critical parameter determining the performance of such configurations, both in atomic ensembles and in the solid state, is the spectral profile of the optical transitions. In the present work, we present a method to fully characterize both the absorptive and dispersive parts of the spectral lines of an absorber inside an optical cavity, including both homogeneous and inhomogeneous broadening effects.  We consider the case when these broadening mechanisms are of similar magnitude.

The effects of introducing dispersive media inside optical cavities are well studied in the context of engineering cavity spectra, for instance cavity linewidth narrowing to give enhanced cavity lifetimes \citep{Soljacic2005} having potential applications in laser stabilisation or high-resolution spectroscopy \citep{Lukin1998}.  Modifications of the spectrum are achieved using dispersion arising from electromagnetically induced transparency \citep{Wang2000}, coherent population trapping \citep{Muller1997} or spectral hole burning \citep{Sabooni2013}.  Likewise the modelling of the susceptibility of warm vapors  has been much studied \citep{Siddons2009}, considering the limiting cases and approximations that may be used.  However, in characterising the susceptibility of atomic or atomic-like systems it is not uncommon to consider approximations such as when the \emph{absorptive} component of the lineshape is dominated by the homogeneous component due to the natural linewidth, for instance in cold atoms \citep{Schultz2008},  or inhomogeneous broadening due to Doppler broadening, where the absorptive component of the lineshape reduces to a Gaussian\footnote{When additionally considering the real (dispersive) part of the susceptability the Lorentzian approximation yields the dispersion function whilst in the Gaussian approximation the dispersion still requires numerical evaluation.}  \citep{Siddons2009}.  It  is noted that using these common approximations helps facilitate the manipulation of the susceptibility, $\chi$, and enable more straightforward computation of the resulting functions.  Such approximations are sufficient when considering either the far-from-resonance or near-resonance limits respectively as compared to the inhomogeneous linewidth  \citep{Siddons2009}.  However, once the atoms are placed in a cavity, it is critical to understand both the dispersion and absorption over a wide frequency range, in order to simultaneously capture the on-resonance absorption of the atoms (and hence access the optical depth and population distribution between the atomic levels), and the resonance conditions of the cavity detuned from the atomic resonances.

Measurements sensitive to both the absorption and dispersion are therefore a useful tool for characterising an atom-filled cavity.  In this case, we gain sensitivity to both of these via the modification of the cavity modes, mapping the dispersion of the intracavity medium into a modulation of the transmitted probe intensity.  Furthermore, the fringe visibility is determined by the intracavity loss which is also determined by the optical depth of the ensemble.  Once set up, this provides a single measurement which simultaneously yields information about the number density, population distribution and broadening mechanisms present of the intracavity atomic medium.

This work describes a full numerical model capturing the absorption and dispersion valid over a wide range of detuning range, in the presence of comparable homogeneous and inhomogeneous broadening. This work was motivated by the emerging requirement to design and calibrate the control systems as part of a wider project based on a warm alkali-vapor filled optical cavity with very specific resonance conditions discussed in \citep{DJS2015}. For such applications high optical depths and coherence times are necessary, where in alkali vapors this often requires both an elevated operating temperature to achieve a high number density, and the addition of a buffer gas in order to maximise the diffusion time of the atoms \citep{Keaveney2014_thesis}.  These together - high temperatures and high buffer gas pressures - can lead to a regime with comparable Doppler and pressure broadening.  With all of these considerations, it was critical to develop a framework with which to understand the electric susceptibility fully.  Nonetheless, the methods and analyses presented here are readily generalised to atoms or atom-like ensembles embedded in cavities.

\section{Model}

\begin{figure}
\center
\includegraphics[width=\linewidth]{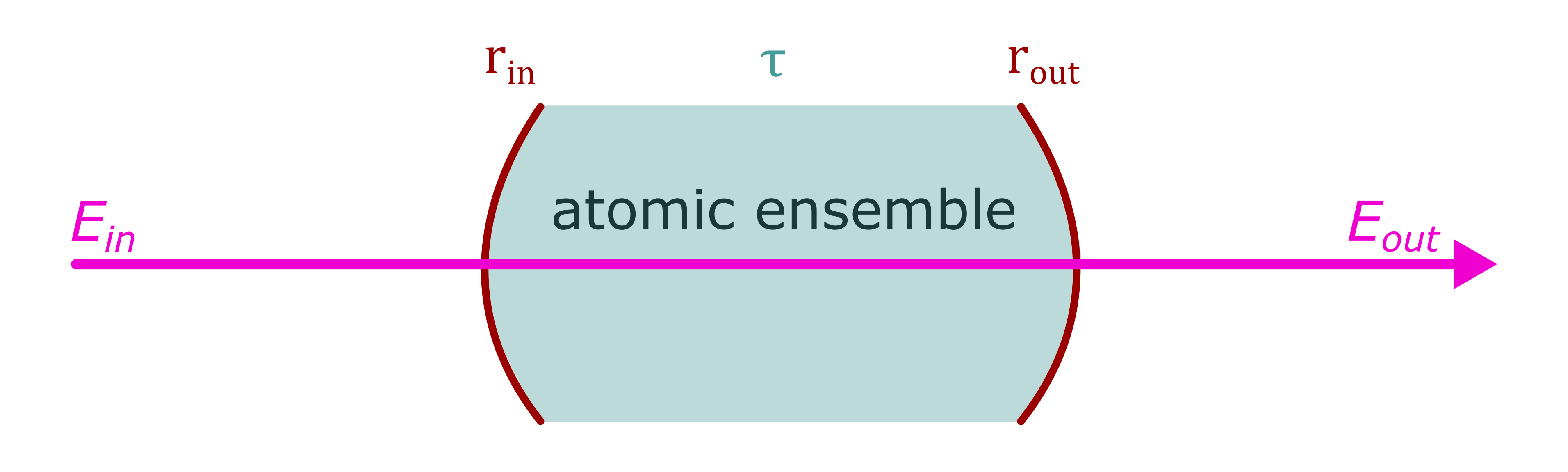}
\caption{(Color online) We consider the transmission response of a cavity contaning some atomic medium, where intracavity ensemble could be warm vapors, cold atoms or solid state systems. \label{fig_simplecav}}
\end{figure}

We begin our analysis by considering a simple Fabry-Perot type cavity containing some atomic ensemble, depicted schematically in figure \ref{fig_simplecav}, the transmission through which can be expressed in the familiar form:
\begin{equation}
E_\mathrm{trans}=t_\mathrm{out}\frac{1}{1-\zeta}\,\,t_\mathrm{in}\,E_\mathrm{in}\label{eqn_CavResp}
\end{equation}
where $\zeta=\tau\,r_\mathrm{in}\,\tau\,r_\mathrm{out}$ is the round trip complex transmission between the outcoupling and incoming mirror, with $\tau$ accounting for the single pass loss and phase accumulated passing through the cavity.

To include the contribution to the atomic ensemble on the cavity response, we begin with the analysis of Siddons \textit{et al.} \citep{Siddons2009} to model the electric susceptibility of the vapor and have based our simulation on the ElecSus computer program \citep{Zentile2015}  published by the same group.  This is then incorporated into our own extended framework to describe the entire atom plus cavity system.

In the expression for the cavity transfer function in equation \ref{eqn_CavResp}, we can account for the frequency-dependent dispersion and absorption of the atomic ensemble\footnote{Additionally, the reflectivities may be polarisation dependent and are not necessarily real scalars, for instance due to stress induced birefringence from vapor cell windows.} in the roundtrip transmission term \mbox{$\zeta=|\zeta\left(\omega\right)|\eulnum^{\im\phi\left(\omega\right)}$}, at laser frequency $\omega$.  For the evaluation of the phase accumulation across the cavity, it is necessary to consider the refractive index integrated over the length $L$ of the ensemble, where the refractive index \mbox{$n=n'+\im n''$} is related to the susceptibility, $\chi$, by
\begin{IEEEeqnarray}{rCl}\label{eqn_chidef}
n&=&\left(1+\chi\right)^{1/2}
\end{IEEEeqnarray}
Separating real ($'$) and imaginary ($''$) parts, we have
\begin{IEEEeqnarray}{rCl}
n'&\approx&1+\frac{\chi'}{2}\\
n''&\approx&\frac{\chi''}{2}
\end{IEEEeqnarray}
for small $\chi$, as is typically the case in our operating regime.
  
The intracavity transmission factor, $\tau$, in equation \ref{eqn_CavResp} can then be written
\begin{IEEEeqnarray}{rCl}
\tau(\omega)&=&\exp\left[  \int_{z=0}^L\ud z\frac{\omega z}{c} \left\{  \frac{-\chi''(\omega,z)}{2} + \im\left(1+\frac{\chi'(\omega,z)}{2}\right)  \right\}  \right]\nonumber\\\label{eqn_tau}
\end{IEEEeqnarray}
and can be substituted into $\zeta$.

The susceptibility can be expressed in terms of the contributions of the dipole moments for each of the atomic transitions. For a given dipole transition between states $\ket{i}$ and $\ket{f}$, the contribution to the susceptibility is of the form \citep{Siddons2008}
\begin{equation}\label{eqn_chiTh}
\chi_{if}=c_{if}^2\frac{N_iD_{if}^2}{\hbar\epsilon_0}s_\lambda(\Delta_{if})
\end{equation}
where $c_{if}$ is the transition strength factor for the given atomic transition, for instance tabulated in \citep{Steck2003} for caesium as considered later in the experimental part of this work, \mbox{$D_{if}=\left\langle i|e\mathbf{r}|f\right\rangle$} is the dipole matrix element, given approximately by \mbox{$D_{if}\approx\left\langle i||e\mathbf{r}||f\right\rangle/3$}, for linear polarisations \citep{Steck2003} assumed in the following analysis, and is the same for all dipole transitions of the D$_2$ line. $N_i$ is the number density of the atoms in state $\ket{i}$  interacting with the light field, and $s_\lambda(\Delta_{if})$ is a shape factor depending on the detuning, \mbox{$\Delta_{if}=\omega-\omega_{if}$}, from the transition \mbox{$i\to f$}, and some set of physical parameters, $\lambda$, as elaborated below.  

With the total susceptibility \mbox{$\chi=\sum_{i,f}\chi_{if}$}, we obtain the complex refractive index of the vapor, giving us the dispersion (real part) and absorption (imaginary part) of the vapor.

For the bare atoms we label the lineshape \mbox{$f_\gamma(\Delta)$}, which is determined solely by $\Delta$ and the natural linewidth of the transition, $\gamma$ (half-width at half-maximum).  For a given transition \mbox{$i\to f$}, the shape factor can be derived \citep{Loudon2000_qtol} as
\begin{IEEEeqnarray*}{rCl}\label{eqn_shapeFac}
f_\gamma(\Delta)&=&\frac{1}{-\Delta+\im\gamma}+\frac{1}{(\Delta+2\omega_{if})+\im\gamma}\\
&\simeq&\frac{-\Delta+\im\gamma}{\Delta^2+\gamma^2}.\yesnumber
\end{IEEEeqnarray*}
In the second line we have taken that \mbox{$\Delta,\,\gamma\ll\omega_{if}$} and make the rotating wave approximation.  In practice, \mbox{$s_\lambda(\Delta)$} must then incorporate the Lorentzian natural linewidth together with contributions from homogeneous and inhomogeneous broadening mechanisms.  In atomic vapor systems these may include collisional broadening and Doppler broadening; or in solid state ensembles similar broadening results from phonons and local charge fluctuations or random strain distribution.

Firstly considering homogeneous broadening,  which in general contributes an additional Lorentzian component to the natural linewidth by adding an additional decay rate to the natural decay rate of the excited state, identically to all emitters.  Here, we consider the specific case of collisional broadening in caesium vapor, where the contribution from pressure broadening and shift of caesium with various buffer gasses is well studied in the literature, see for instance \citep{Pitz2010,Kozlova2011}. Nonetheless, analogous broadening occurs in solid state systems. Collisional broadening can therefore be described by including an additional component to the Lorentzian natural linewidth \citep{Loudon2000_qtol} $f$:
\begin{equation}
\gamma_{\mathrm{P}}=\gamma+\gamma_{\mathrm{coll}}
\end{equation}
where $\gamma$ is the natural linewidth, and \mbox{$\gamma_{\mathrm{coll}}\propto n\bar{v}_{\mathrm{th}}$} is the contribution from collision induced broadening, where $n$ is the number density and $\bar{v}_{\mathrm{th}}$ is the thermal velocity of the atoms and the constant of proportionality relates to their collisional cross section. For experiments with relatively low atomic densities of caesium, the pressure broadening is dominated by the buffer gas.  Similarly, the buffer gas induces a shift in the energy of the transitions, where the shift is also linear in the pressure.

Inhomogeneous broadening due to the varying local environment of each emitter has the effect of convolving the refractive index over the probability distribution of shifted resonance frequencies.  In the case of Doppler broadening this means convolving over the thermal velocity distribution of the atoms, due to the Doppler shifted frequency \mbox{$\omega'=\omega(1-v_z/c)=\omega-\delta_v$}, seen by an atom with a velocity component $v_z$ parallel to the direction of beam propagation.  The resulting shape, as a function of detuning is then given by the convolution
\begin{IEEEeqnarray}{rCl}
s_{\gamma,\sigma}(\Delta)&=&[f_\gamma\ast g_\sigma](\Delta)\\
&=&\int_{-\infty}^{\infty}f_\gamma(\Delta-\delta_v)g_\sigma(\delta_v)\ud \delta_v
\end{IEEEeqnarray}
where \mbox{$g_{\sigma}(\delta_v)$} is the Gaussian distribution of frequency shift due to velocity in one dimension
\begin{IEEEeqnarray}{rCl}
g_\sigma(\delta_v)&=&\frac{1}{\sigma\sqrt{2\pi}}\exp\left[\frac{-1}{2}\left(\frac{\delta_v}{\sigma}\right)^2\right],\label{eqn_gaussNotation}
\end{IEEEeqnarray}
and
\begin{IEEEeqnarray}{rCl}
\delta_v&=&\frac{\omega v_z}{c}\\
\sigma&=&\frac{\omega}{c}\sqrt{\frac{k_BT}{m}}.
\end{IEEEeqnarray}
The resulting shape function is related to the Faddeeva function (if $f_\gamma$ is taken before making the rotating wave approximation in equation \ref{eqn_shapeFac}) \citep{Siddons2009}, and arises in a number of physical systems; nevertheless, it is difficult to handle numerically and has therefore attracted an amount of research over the years into finding either analytic approximations (e.g. \citep{Thompson1986}) or more recently developing efficient algorithms (e.g. \citep{Wells1999,SGJohnson2012}).  The absorptive component results in the well known Voigt function.  On the other hand, atomic \textit{dispersion} is in general more difficult to measure directly, with experiments requiring actively stabilised interferometry, for instance \citep{Wicht2000}.  Our approach enables access to the dispersive part of the lineshape function through the phase accummulated by the probe light on each round trip of the cavity.

Assuming that the natural linewidth $\gamma$ (and likewise the pressure width $\gamma_p$, and Doppler width $\sigma$) are identical for each transition, as in \citep{Zentile2015}, this shape function can then be inserted into equation \ref{eqn_chiTh} to obtain the total refractive index of the medium, and thus find the resulting cavity transmission function.  

Having obtained the full expression for the cavity response including the contribution from the vapor, we note that the strength of light matter interaction is often discussed in terms of the optical depth, $d$, of the matter ensemble, where $d$ is defined by the on resonance absorption of the medium: \mbox{$I_{\mathrm{out}}=I_{\mathrm{in}}\exp\left(-2d\right)$}. 

Then for the bare atoms (i.e. neglecting any Doppler or pressure effects) perfectly initialised into the ground state $i$, from equation \ref{eqn_shapeFac} on resonance we have that \mbox{$s(0)=\im/\gamma$}, so $d$ can be expressed as \citep{Reim2011_thesis}
\begin{IEEEeqnarray*}{rCl}
d&=&\frac{1}{2}\frac{\omega_0L}{c}\frac{ND_{if}^2}{\hbar\epsilon_0}\frac{1}{\gamma},\yesnumber
\end{IEEEeqnarray*}
where $N$ is the number density of caesium atoms interacting with the light field, in some length $L$.  With this we can rewrite the susceptibility as
\begin{IEEEeqnarray*}{rCl}
\chi&=&\frac{2d\gamma}{\omega_0L/c}s_\lambda(\Delta).\yesnumber 
\end{IEEEeqnarray*}
Noting that far away from resonance, the Lorentzian character of the shape profile, \mbox{$s(\Delta)$}, is dominant \citep{Siddons2009}, in the limit of large detuning the absorption and dispersion have the functional dependance
\begin{IEEEeqnarray}{rCl}
\lim_{\Delta\gg\gamma,\sigma}\left(\frac{\omega L}{c}n'\right)&=&\frac{\omega L}{c}-\frac{\omega}{\omega_0}\frac{d\gamma}{\Delta}\\
\lim_{\Delta\gg\gamma,\sigma}\left(\frac{\omega L}{c}n''\right)&=&\frac{\omega}{\omega_0}\frac{d\gamma^2}{\Delta^2}.
\end{IEEEeqnarray}

Noting that the dispersive part ($n'$) of the complex refractive index decays slower \mbox{($1/\Delta$)} away from resonance, this means that measurements sensitive to the dispersion yield information about the atomic population over a wider frequency range than measuring only the absorption (which scales as \mbox{$1/\Delta^2$)}, as exploited in reference \citep{Sprague2014}. This affects the frequencies of the cavity resonances and therefore provides both a means to measure the atomic populations and linewidths as well as to control the position of the full cavity transmission frequencies. 

With this description, then,  the intra cavity field and transmission can be obtained from \ref{eqn_CavResp} (via equation \ref{eqn_tau}), as a function of physical parameters: the temperature of the vapor (broadening), the caesium reservoir temperature (number density), population distribution between the ground states (optical pumping efficiency) and the cavity reflectivities.  \\

\section{Experiment}

\begin{figure}[!t]
\center
\subfloat[\label{fig_cavschem}]{
\includegraphics[width=.8\linewidth]{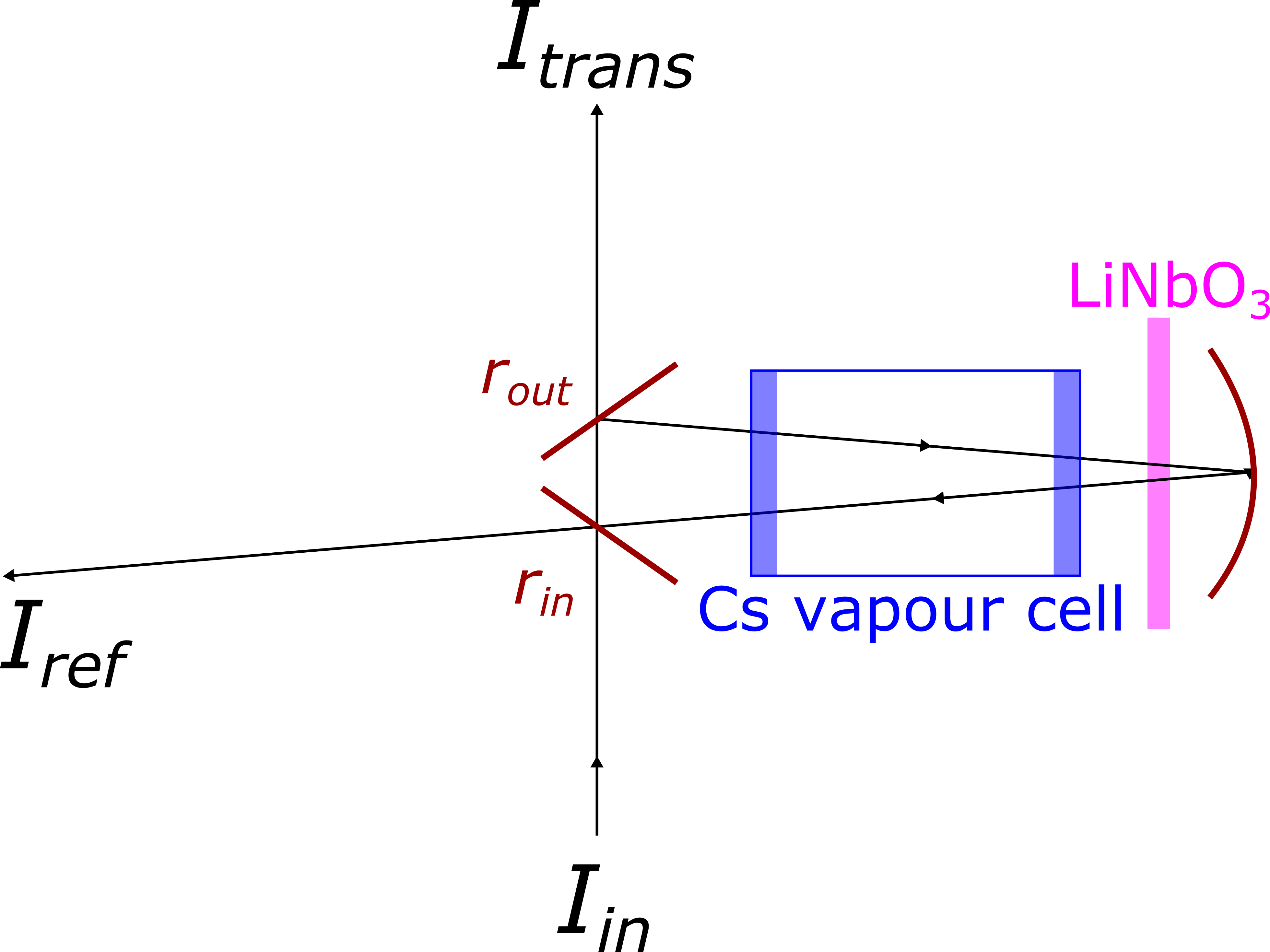}}\\
\subfloat[\label{fig_lambdaPump}]{
\includegraphics[width=.8\linewidth]{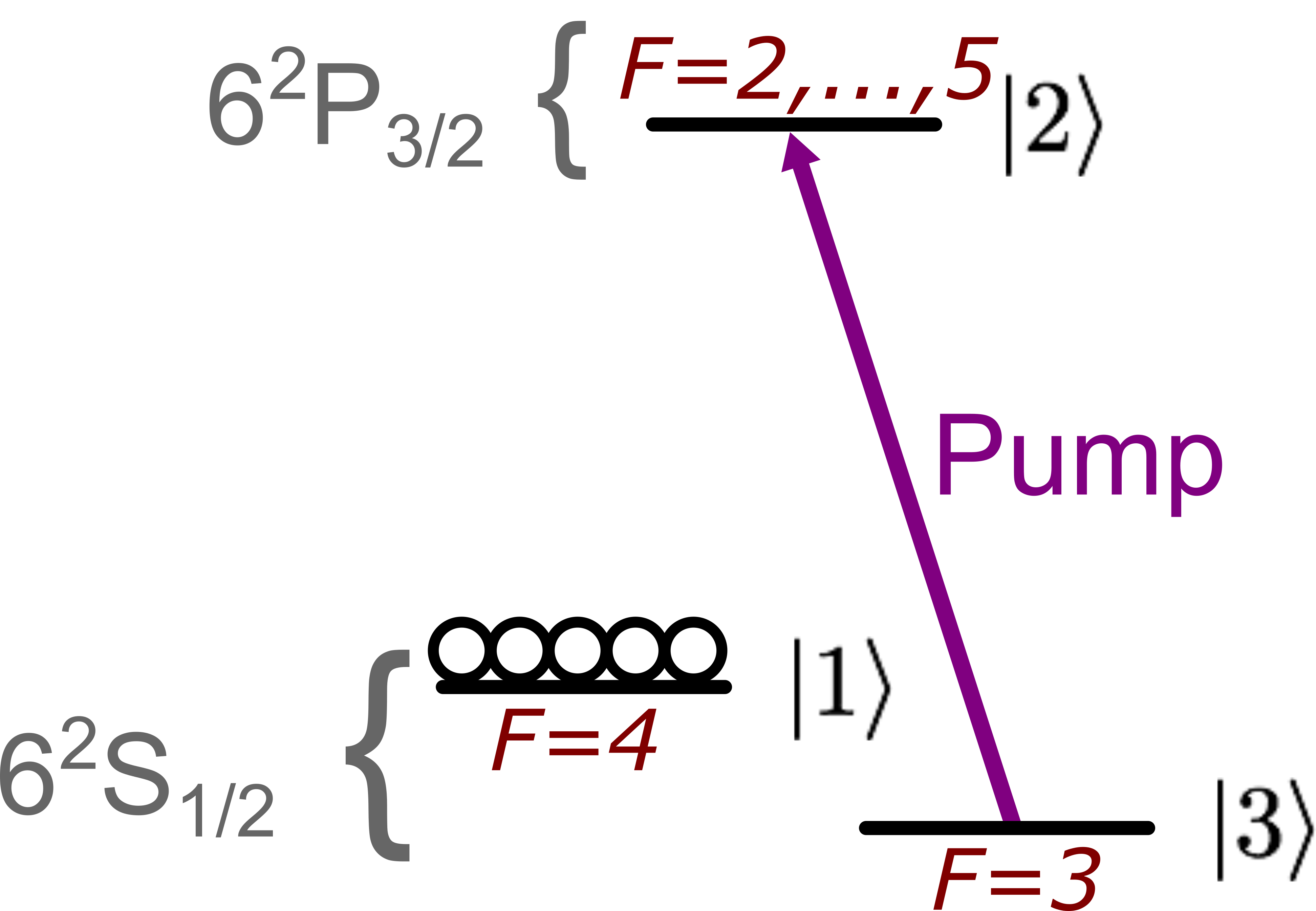}}
\caption{(Color online) \ref{fig_cavschem} Schematic of the ring cavity design experimentally implemented; \ref{fig_lambdaPump} Energy level $\Lambda$ configuration of the caesium D$_2$ transition.  The ensemble is initialised into the \mbox{$F=4$} hyperfine ground state via optical pumping. }
\end{figure}

We now apply the general model developed in the previous section to our particular experimental arrangement.  The system analysed in the present work is comprised of a ring cavity, passing through caesium-filled vapor cell twice each roundtrip as illustrated in figure \ref{fig_cavschem}.  The vapor cell (Precision Glassblowing) of length 12mm and diameter 10mm also contained \mbox{10 Torr} of neon buffer gas.  To control the number density of caesium in the vapor phase, the cell was heated by a coil of quad-twisted wire (Lakeshore QT-32) wrapped around the cell, with thermistors between the wire and the cell surface connected to a temperature controller, controlling the current through the coil.  The central upper region of the cell was cooled by a cold finger driven by a Peltier element in order to prevent caesium deposition on the cell windows.

Note that the ring cavity configuration doubles the effective length of caesium compared to free space.  Another motivation for using a ring cavity for light matter interaction experiments using warm vapor is to avoid the formation of a  standing wave as would form in a linear cavity where the intensity modulation over length scales of a wavelength would otherwise limit the coherence lifetime due to diffusion of the atoms.

A schematic of the energy level structure of the D$_2$ line of caesium is shown in  figure \ref{fig_lambdaPump}, where the ensemble of atoms are optically pumped into one of the hyperfine groundstate levels.  In the present analysis we consider the case of zero magnetic field for simplicity, and so we neglect consideration of the hyperfine sublevels.  Linear polarisation is used for both the optical pump and signal light.  To account for the effect of imperfect pumping of the atoms, however, when evaluating the susceptability we therefore sum the contributions of each transition to the overall susceptibility with their appropriate detunings and weighted according to their populations.  The population distribution is defined in terms of the atomic polarisation, \mbox{$w=p_1-p_3$}$\in\mbox{[-1,1]}$ where $p_i$ are the population fractions of the hyperfine ground state levels $i$.  Subsequently we also use $w$ as the metric for the optical pumping efficiency.

For the purposes of \citep{DJS2015} the cavity is configured with a free spectral range of \mbox{$\Delta\omega\sim7\,\mathrm{GHz}$} in the absence of the vapor, with an acceptance bandwidth \mbox{$\delta\omega\sim1\,\mathrm{GHz}$}. This configuration is chosen in order to meet the desired resonance condition \mbox{$2\Delta_\mathrm{G}=(m+1/2)\Delta\omega$} where \mbox{$\Delta_\mathrm{G}=9.2\,\mathrm{GHz}$} is the hyperfine ground state splitting of caesium and $m$ is some integer.  In this case, \mbox{$m=2$} is chosen giving a roundtrip cavity length \mbox{$\sim40\,\mathrm{mm}$}.

Additionally, a \mbox{$200\,\mu\mathrm{m}$} \mbox{LiNb0$_3$} crystal  is included in the cavity to provide controllable birefringence in order to independently manipulate the relative optical path lengths for orthogonal polarisations inside the cavity, via temperature tuning.

With this picture, equation \ref{eqn_CavResp} is modified slightly to become:
\begin{IEEEeqnarray}{rCl}
\frac{E_\mathrm{trans}}{E_\mathrm{in}}&=&\frac{t_\mathrm{in}t_\mathrm{out}\eulnum^{\im\delta}}{1-\zeta}\label{eqn_CavResp_ring}
\end{IEEEeqnarray}
where now $\delta$ is the component of the phase acquired between the input and output mirrors where we assume no loss. From this, we can then extract expressions for the free spectral range (FSR, $\Delta\omega$), bandwidth ($\delta\omega$), and finesse ($F$).

Having chosen the FSR (\mbox{$\Delta\omega=2\pi c/L$}) in order to meet the specific multi-resonances conditions above, appropriate reflectivities can be chosen such that the cavity bandwidth matches that of our input signal, with a FWHM of \mbox{1.2 GHz}, via $F$
\begin{IEEEeqnarray*}{rCcCl}
F&=&\frac{\Delta\omega}{\delta\omega}
&=&\frac{\pi}{2\arcsin\left(\frac{1-\zeta}{2\sqrt{\zeta}}\right)}.\yesnumber
\end{IEEEeqnarray*}\\

As before, separating the absorptive and dispersive parts, writing \mbox{$\zeta=|\zeta|\eulnum^{\im\phi}$}, then gives
\begin{IEEEeqnarray}{rCl}
|\zeta(\omega)|&=&\zeta_0\exp\left[-\frac{\omega}{c}n''(\omega)L_{\mathrm{Cs}}\right]\label{eqn_CavLoss}\\
\phi(\omega)&=&\frac{\omega}{c}\left[n'(\omega)L_{\mathrm{Cs}}+L_{\mathrm{out}}\right]\label{eqn_CavPhase},
\end{IEEEeqnarray}
where $L_{\mathrm{Cs}}$ is the total distance of caesium travelled through in one round trip, $L_{\mathrm{out}}$ is the additional path length outside of the vapor cell, and $n$ is the complex refractive index of the caesium.  Note that we have assumed that the contribution to $\zeta$ from the mirrors is purely real and is included in $\zeta_0$, along with any other intracavity losses. 

For the vapor cell used in this experiment, estimates for the pressure induced broadening and shift of the caesium transitions  due to the buffer gas, at a given temperature was made using the results presented in reference \citep{Pitz2010} in tables II and III respectively.  Here the self contribution to pressure broadening from caesium is negligible due to the low densities in our operating regime \mbox{$\sim10^{12}\,\mathrm{cm}^{-3}$} \citep{Weller2011}.   The total number density of caesium atoms in the vapor phase is estimated from the temperature of the caesium reservoir at the coldest point in the cell, using the expression for the vapor pressure, $P_\mathrm{v}(T)$, given in equation 1 of \citep{Steck2003}, via the ideal gas law \mbox{$n_\mathrm{v}(T)=P_\mathrm{v}/k_\mathrm{B}T$}.  The reflectivities, transmissions and losses for the cavity optics were also characterised experimentally.
 
The frequency of a probe laser was scanned and the transmission through the cavity in the vicinity of the D$_2$ transitions measured. Low probe powers  \mbox{($\sim\mu\mathrm{W}$)} are used so as not to affect the  population distribution during the measurement. Comparison with this model allows us to obtain an estimate of the physical parameters: critically the optical depth and the pumping efficiency which are key figures of merit for the memory interaction in \citep{DJS2015}.  \\

\subsection{Experimental Details}

The stability of the cavity is critical for these measurements: in particular, the length and temperature of the caesium cell were actively stabilised.  The cavity length was locked using a variant of the H\"{a}nsch-Couillaud method \citep{Hansch1980}, where the polarisation rotation of a HeNe laser generates a length dependent error signal which is monitored by a microcontroller (Arduino Due) that provides both fast (kHz) and slow (Hz) feedback to a piezo-electric stack on which the concave cavity mirror is mounted. The cavity lock remains stable to $\pm50\,\mathrm{MHz}$ for a duration of hours.  Similarly, the cavity response is strongly dependent on the optical depth of the caesium, and hence temperature stability is also critical.  The temperature control infrastructure remained stable to \mbox{$\pm0.1\,\mathrm{K}$} over the same timescale.

The scans were conducted with the cavity locked to maximise transmission at \mbox{$\Delta=+6\,\mathrm{GHz}$} from the \mbox{$F=3$} transition and a CW diode laser (Toptica DL Pro, \mbox{$<$1 MHz} linewidth) locked to the $\ket{6^2S_{1/2};F=3}\to\ket{6^2P_{3/2};F=3}$, $\ket{6^2S_{1/2};F=3}\to\ket{6^2P_{3/2};F=4}$ crossover transition to optically pump the ensemble into the \mbox{$F=4$} hyperfine ground state. The modelocked TiSa (Spectra Physics Tsunami \mbox{$\sim1\,\mathrm{GHz}$} linewidth) was then scanned across both of the D$_2$ resonances, and the cavity transmission recorded (Fig. \ref{fig_scanEg}), where the mode-hop free tuning range limited the scans to a range of \mbox{$\sim$30 GHz}.  See \citep{DJS2015} for further details.  
 
Note that, to fully model the cavity response observed with a given probe laser, the cavity response model must also be convolved with the frequency profile of the probe: which in this case we take to be a Gaussian \mbox{$g_{\sigma_\mathrm{p}}(\omega-\omega_0)$}, with FWHM (in intensity) of \mbox{$0.5\,\mathrm{GHz}$}.\\

\subsection{Numerics}
 
The resulting function to which we fit the data is given in full by equations \ref{eqn_fullTransmission}-\ref{eqn_linshape_fin} below.  Together, these define the cost function optimised when fitting the data.

\begin{widetext}
\begin{equation}
\frac{I_\mathrm{out}(\omega,\{T_k\})}{I_\mathrm{In}}\approx
A+B
g_{\sigma_\mathrm{p}}(\omega)\ast\left\vert
\exp\left[\frac{\im\omega L_\mathrm{out}}{c}+\im\phi\right]\left(
1+\left|\zeta_0\right|\exp\left[
\frac{\im\omega L_\mathrm{Cs}}{c}\left(1+\frac{1}{2}\sum_{ij}\chi_{ij}(\omega,\{T_k\})\right)+\frac{\im\omega L_\mathrm{out}}{c}+\im\phi
\right]
\right)^{-1}
\right\vert^2,\label{eqn_fullTransmission}\\
\end{equation}
\end{widetext}

where
\begin{IEEEeqnarray*}{rCl}
\chi_{ij}(\omega,\{T_k\})&=&c_{ij}^2\frac{p_i n_\mathrm{v}(T_\mathrm{res}) D_{ij}^2}{\hbar\epsilon_0} s\big[\omega-\omega_{ij}-\delta_\mathrm{coll}(T_\mathrm{P}),\{T_k\}\big]\\
s(\Delta,\{T_k\})&=&\left[ f_{\gamma+\gamma_\mathrm{coll}(T_\mathrm{P})}\ast g_{\sigma_\mathrm{D}(T_\mathrm{D})} \right] (\Delta).\label{eqn_linshape_fin}\yesnumber\\
\end{IEEEeqnarray*}

The coefficients $A$ and $B$ are experimental parameters determined by the background and overall scaling, but the remaining constants can be used to extract the temperature via the Doppler width $\sigma_\mathrm{D}(T_\mathrm{D})$, the pressure broadening $\gamma_\mathrm{coll}(T_\mathrm{P})$ and shift $\delta_\mathrm{coll}(T_\mathrm{P})$, and the number $n_\mathrm{v}(T_\mathrm{res})$ of atoms, and populations $p_i$ of the hyperfine ground state levels $i$.

In the preceeding expressions, we have distinguished three effective temperatures $\{T_k\}$, where the indices \mbox{$k\in\{$`res',`P',`D'$\}$} refer to the temperatures corresponding to the caesium reservoir which determines the number density, pressure broadening and shift, and Doppler broadening.  We permit these to vary independently as they are not necessarily equal \citep{Zentile2015} due to the non-uniform temperature within the cell and to account for the limited accuracy of the formulae for number density and pressure width.  Additionally, we include a correction phase \mbox{$\phi\in[-\pi,\pi]$} to the external cavity length $L_\mathrm{out}$, where $\phi$ is determined by the setpoint of the cavity lock.

The functions $g_\sigma(\omega)$ and $f_\gamma(\omega)$ are defined as in equations \ref{eqn_gaussNotation} and \ref{eqn_shapeFac}.  The convolution  in equation \ref{eqn_linshape_fin} combines the homogeneous and inhomogeneous broadening terms, and in \ref{eqn_fullTransmission} accounts for the bandwidth of the probe laser.We note that as an alternative to performing the convolution of $f$ and $g$ explicitly, there exist algorithms for evaluating the Faddeeva function, see \citep{SGJohnson2012}.   This method was also implemented and compared with the output from performing the convolution; however, both methods performed comparably, with the numerics for the convolution being slightly faster in our implementation.

Due to the number of parameters describing the system in equations \ref{eqn_fullTransmission}-\ref{eqn_linshape_fin}, this problem is a difficult one to search numerically as the natural periodicity of the function results in many local minima, and each evaluation of the profile is itself intensive due to the two nested convolutions or the calls to the Faddeeva package.  In order to make the problem more tractable two ``one off'' calibration measurements were performed aiming either to constrain variables or to provide favourable starting estimates for the fitting routine.

Firstly,  in order to relate the measured surface temperature of the cell to the effective temperatures of the vapor, the reservoir temperature and buffer gas effects were estimated at a range of temperatures by measuring the absorption spectrum of the caesium in the absence of the cavity. Fitting the absorption spectrum is less numerically intensive than for the whole cavity spectrum and so we readily obtain values for the reservoir temperature and effective temperature of the motional broadening at a given heater wire current.  As indicated previously, treating these independently allows for differential temperature accross the cell and uncertainties in the expressions used to define the widths and number density.   We note that for the temperature range investigated, it was observed allowing the effective temperatures ($T_\mathrm{res}$, $T_\mathrm{P}$, $T_\mathrm{D}$) to vary independently yields a better fit than defining a single temperature parameter. At lower temperatures it was observed that the obtained effective motional temperature (Doppler and pressure) is consistently greater than the reservoir temperature (\mbox{$T_\mathrm{res}<(T_\mathrm{P}, T_\mathrm{D})$}), whilst at higher temperatures both approaches gave comparable results (\mbox{$T_\mathrm{res}\approx T_\mathrm{P}\approx T_\mathrm{D}$}).  The crossover between these regimes occurs at \mbox{$\sim60^\circ\mathrm{C}$}.

Secondly, with the cavity constructed around the cell, the roundtrip distance was constrained by measuring the cavity response far from the atomic resonance where dispersion is negligible.  This constrains the cavity length (\mbox{$L_\mathrm{out}+L_\mathrm{Cs}$}) to within a wavelength, and fixes the intracavity loss $\zeta_0$.  In the subsequent measurements, the total cavity length was then fixed by the correction phase $\phi$ such that the cavity is locked to maximise transmission at a detuning of \mbox{$+6\,\mathrm{GHz}$}.

\begin{figure}[!tp]
\center
\subfloat[Pumping efficiency\label{subfig_eta}]{
\includegraphics[width=\linewidth]{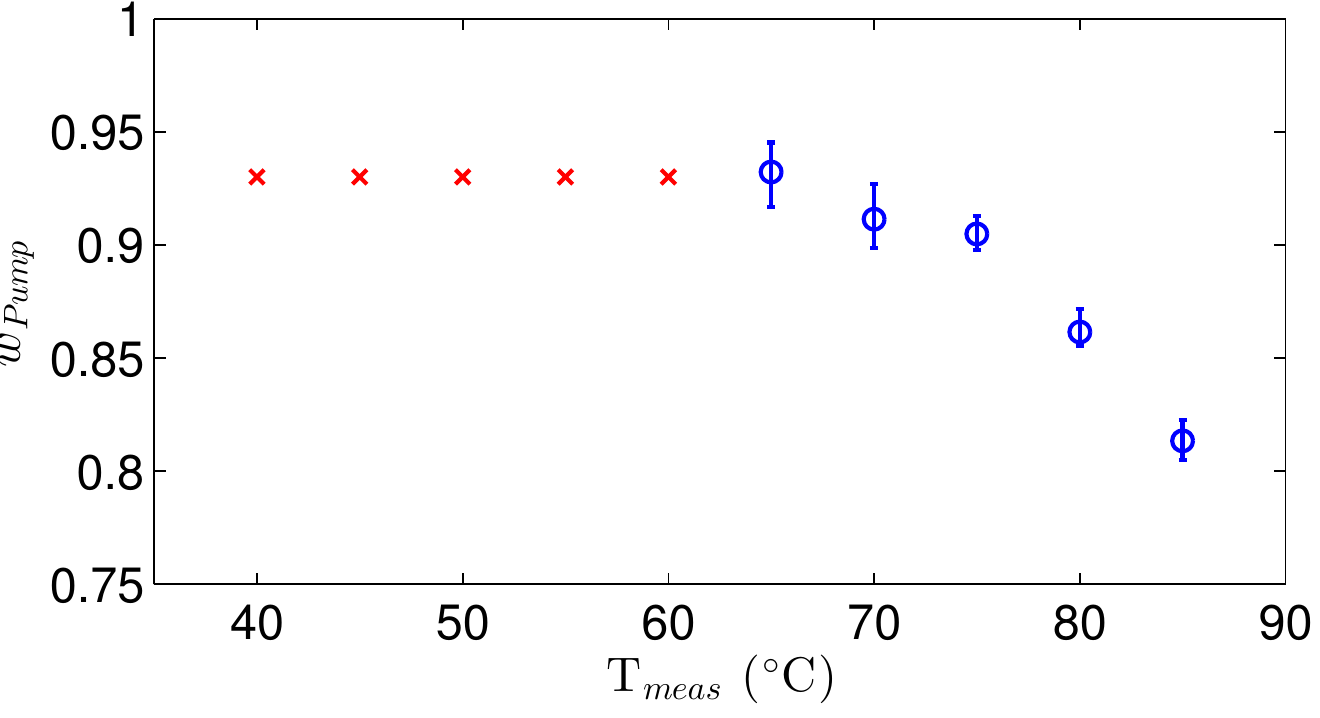}}\\
\subfloat[Caesium number density\label{subfig_num}]{
\includegraphics[width=\linewidth]{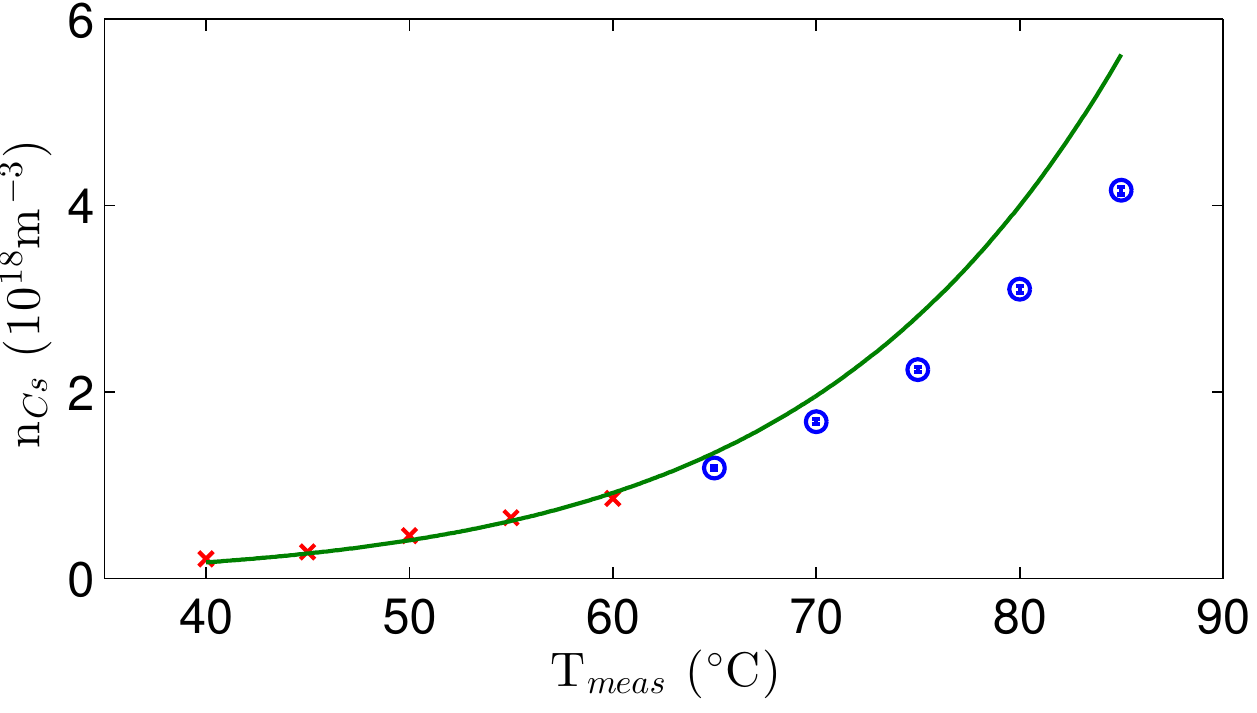}}
\caption{(Color online) Extracted parameters as a function of measured temperature.  Blue circles indicate values obtained using the automated fitting routine, and red crosses those obtained using manual fits.\ref{subfig_eta} optical pumping efficiency:  the fall in pumping efficiency seen at \mbox{$60^\circ\mathrm{C}$} indicates the temperature below which the automated fitting routine no longer converged.  The plateau in polarisation as the temperature is reduced (instead of increasing to unity), observed in these data likely arises from a combination of imperfect mode overlap and limited pumping light entering the cavity; \ref{subfig_num} caesium number density: the green line gives the number density expected if the caesium reservoir were held at the temperature measured at the cell surface.\label{fig_paramExtract}}
\end{figure}

\begin{figure*}[!th]
\center
\subfloat[Manual fit at $40^\circ\,\mathrm{C}$ \label{subfig_pf10}]{
\includegraphics[width=.48\linewidth]{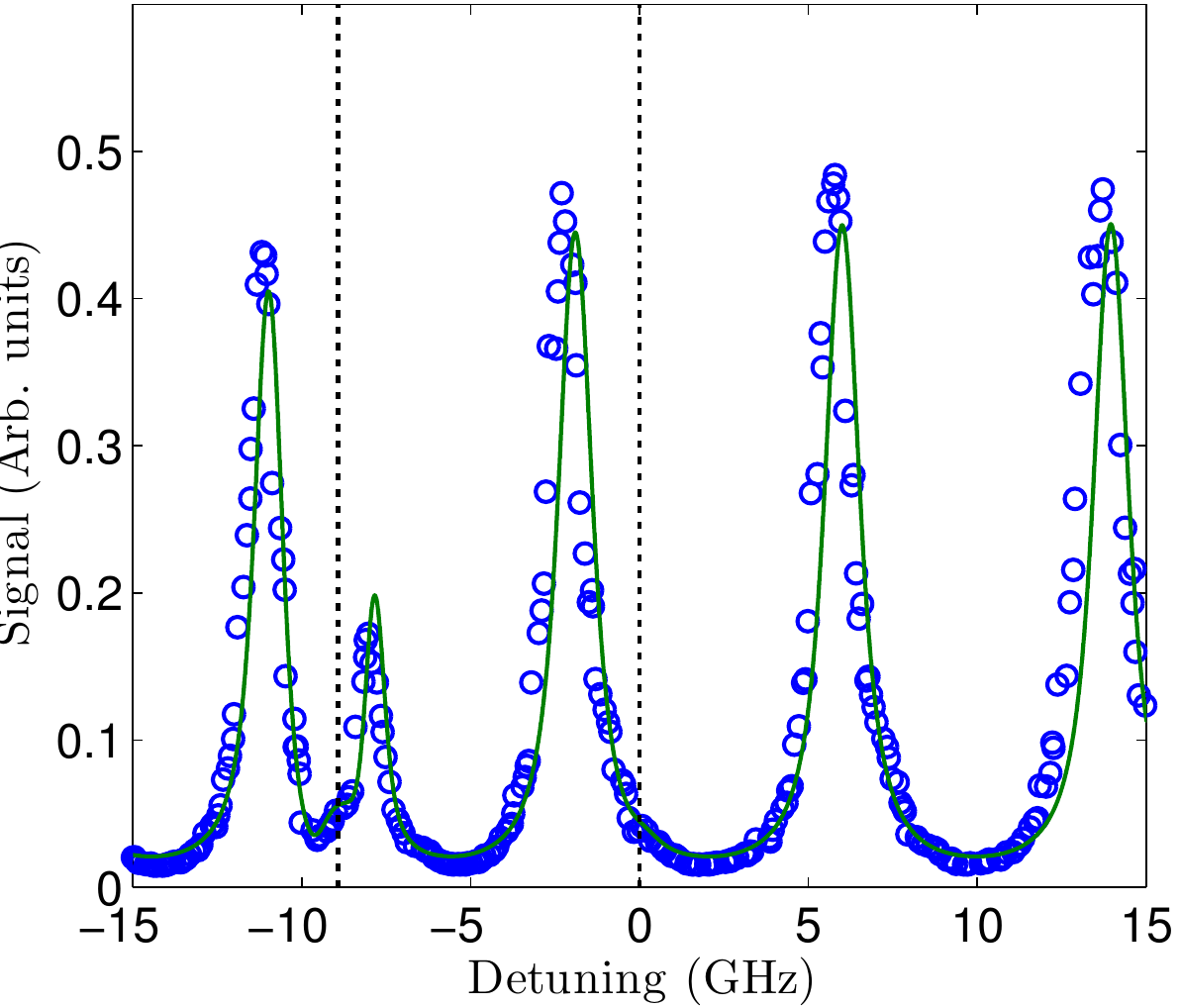}}\hfill
\subfloat[Automatic fit at $85^\circ\,\mathrm{C}$\label{subfig_af1}]{
\includegraphics[width=.48\linewidth]{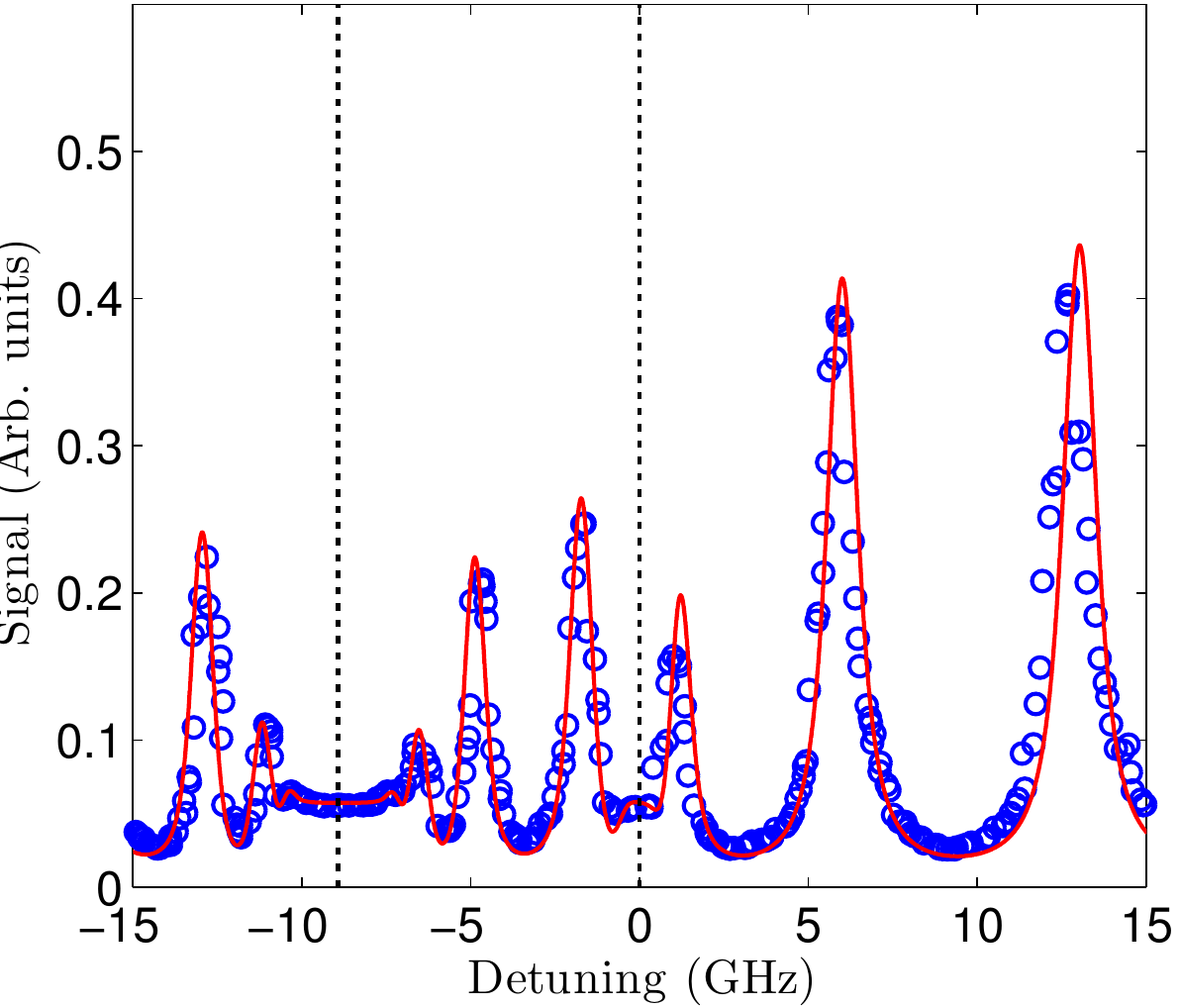}}\\
\subfloat[Comparison between response for filled and empty cavity.]{
\includegraphics[width=\linewidth]{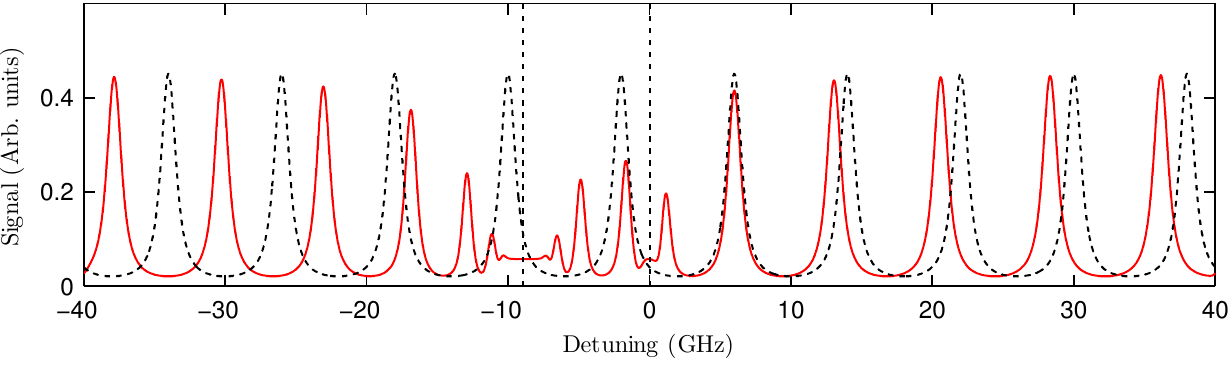}}
\caption{(Color online) Example of (a) manual search result at \mbox{$40^\circ\mathrm{C}$} and (b) automatic fit at \mbox{$85^\circ\mathrm{C}$}.  The dashed lines indicate the frequencies corresponding to the centre of the two hyperfine resonances, \mbox{$F=4$} at \mbox{$-9.2\,\mathrm{GHz}$} and \mbox{$F=3$} at 0 detuning.  Circles indicate experimental data, and the lines the theoretical response for manually fitted paramaters. (c) Comparison of modelled cavity response from automatic fit at \mbox{$85^\circ\,\mathrm{C}$} - (as in (b), red) with cavity response with no atoms present (black, dashed).  \label{fig_scanEg}}
\end{figure*}

Based on the above, one approach for extracting parameters was to constrain the pressure and Doppler temperatures to equal that of the reservoir so as to reduce the size of the parameter space to be searched.  The pumping efficiency and reservoir temperature, which together determine the optical depth,  extracted with this method are plotted in figure \ref{fig_paramExtract}. As observed in the measurement of the absorption spectrum, at higher temperatures \mbox{($\gtrsim60^\circ\mathrm{C}$)} the assumption of a single temperature seems to reliably yield good fits, whereas at lower temperatures the fitting method became less robust.   A typical fit obtained using this method is shown in figure \ref{subfig_af1} for measurements taken at the maximum operating temperature. The Peltier element was driven at a constant power and therefore the temperature differential across the cell varies with the current driving the heater wire temperature which was used to set the reservoir temperature.  This may explain why the equal temperature approximation, required for the numerics to converge, breaks down at lower temperatures, and in this regime we rely upon manual searches. 

Manual searches were performed by iterating over the Doppler and pressure widths and the reservoir temperature perturbed  over some physically reasonable range from those obtained in fitting the absorption spectra at the same heater wire current, allowing for variation in environmental conditions.  The cost function was then optimised over the remaining unconstrained parameters, namely the pumping efficiency ($w$), background intensity on the detector ($A$) and the overall scaling of the photodiode measurements ($B$). For this approach we generate the modelled cavity response drawing parameters from a 6 dimensional array containing the perturbed parameters ($T_\mathrm{res}$, $\sigma_\mathrm{D}$ and  $\gamma_\mathrm{P}$), and a range of estimates for $A$, $B$ and $w$.  From this, we then identify the set of parameters within the array yielding a minimum in residuals.  Typical fits obtained using this method shown in figure \ref{subfig_pf10} for measurements taken at the minimum operating temperature.  We note that at a measured temperature of \mbox{$85^\circ\mathrm{C}$} at the surface of the cell, we obtain the approximate values given in table~\ref{tab_ManParams}, which are comparable to those obtained using the automated fits with a single temperature parameter.  

\begin{table}
\begin{center}
\begin{tabular}{|c c c|} \hline
Parameter & Manual & Automated \\ \hline
$T_\mathrm{Res}\,/\,^\circ\mathrm{C}$ & 80.8 & 80.6\\ 
$\gamma_\mathrm{P}\,/\,\mathrm{MHz}$ & 41 & 65\\
$\sigma_\mathrm{D}\,/\,\mathrm{MHz}$ & 160 & 175\\ 
$\eta$ & 0.80 & 0.81\\ \hline
\end{tabular}
\end{center}
\caption{(Color online) Parameters extracted from the two fitting routines for data obtained at a measured temperature of \mbox{$85^\circ\mathrm{C}$} at the cell surface.\label{tab_ManParams}}
\end{table}

To benchmark the performance of the fitting routine, we investigate the pumping efficiency and number density as functions of temperature, both of which are well studied \citep{Rosenberry2007,Steck2003}.  The pumping efficiency is seen to roll off at \mbox{$\sim70^\circ\mathrm{C}$}, at which point increasing the temperature, and hence density of Cs, leads to light trapping impeding the ability to optically pump efficiently \citep{Rosenberry2007}.  The extracted reservoir temperature is in good agreement with the recorded temperature at the surface of the cell.  Observation of this known behaviour confirms our numerical modelling.  Error bars are generated for the automated fitting routine using the Monte-Carlo method over 100 trials.\\

\begin{figure*}[!t]
\center
\includegraphics[width=.8\textwidth]{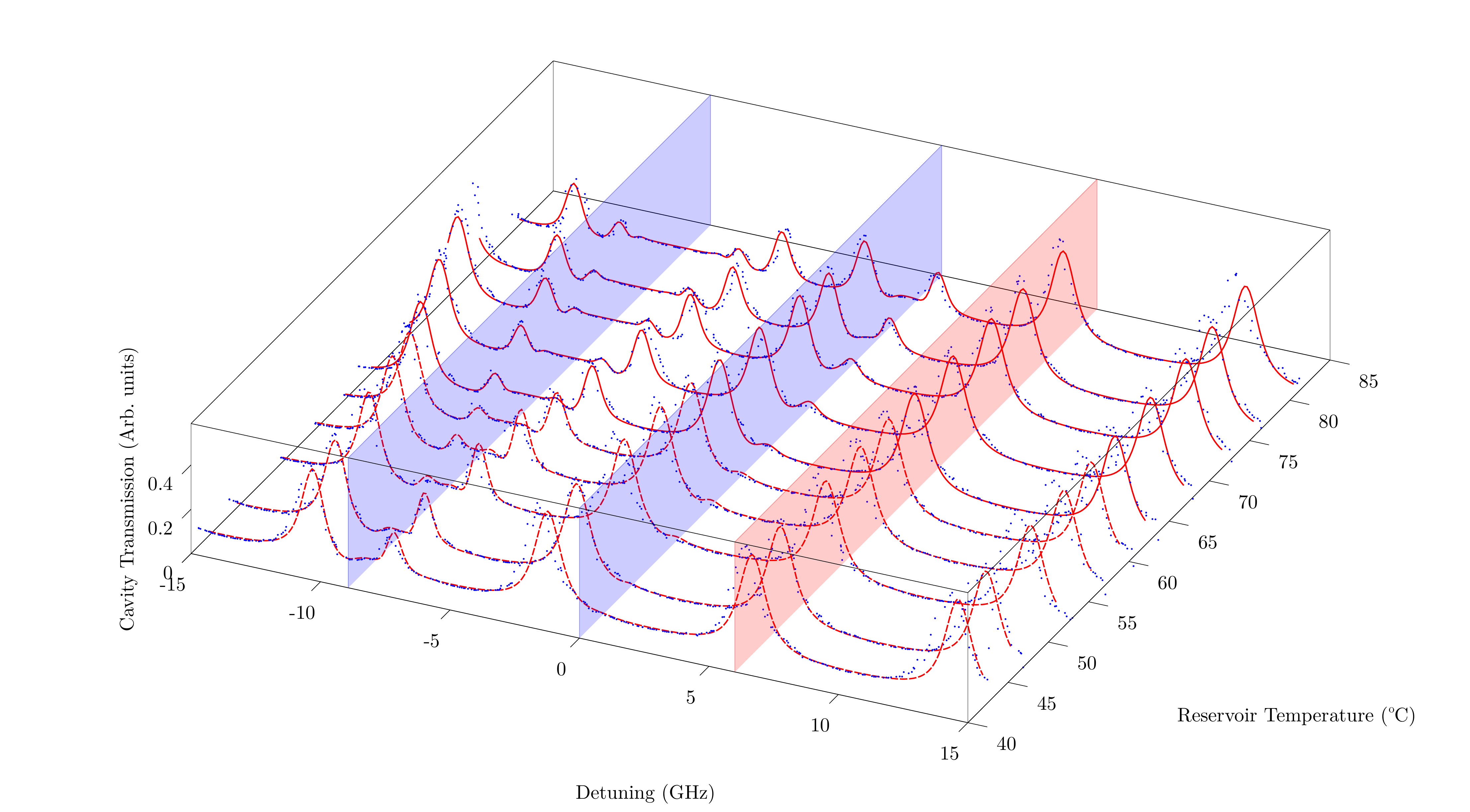}
\caption{(Color online) Cavity transmission, locked to resonance at \mbox{$6\,\mathrm{GHz}$} blue detuned from the \mbox{$F=3$} resonance. Showing our model, from fitted parameters obtained from measurements at \mbox{$5^\circ\mathrm{C}$} intervals.  Blue surfaces indicate the frequencies of the two D$_2$ resonances, and the red surface the position of the cavity lock.  Solid red lines indicate the model obtained using a fitting routine, red dashed lines show the manually obtained models.\label{fig_responseLandscape}}
\end{figure*}

For the purposes of \citep{DJS2015}, as in other experiments of atomic-like systems in cavities, control of the optical depth, pumping efficiency and cavity resonance frequencies are critical.  In the present work this was demonstrated, and  the variation of these key parameters was investigated as a function of temperature with caesium vapor in a cavity.  These scans were recorded at \mbox{$5\,\mathrm{K}$} intervals, and are shown in figure \ref{fig_responseLandscape}.  The response of the cavity, falling pumping efficiency and increasing number density of caesium with temperature seen in the data are captured well by the model.

\section{Conclusions}

This framework provides a basis for characterisation of the susceptibility of atom-like systems in a cavity, where the frequency stability of the cavity is used as a reference to obtain information about dispersion.  Access to both the absorption and dispersion over a wide range of frequencies (near- and far-detuned) provides a diagnostic tool to confirm useful operation parameters (in this case pumping efficiency, optical depth and cavity resonance conditions) \textit{in situ} via a simple measurement of the cavity transmission.  This is additionally useful since, whilst measurements of the pumping efficiency outside the cavity via absorption are straightforward, the cavity will modify the ability to optically pump the ensemble.  If properly handled, measurement of the cavity transmission can provide a single shot measurement yielding a number of critical operating parameters.

In addition, this tool can therefore be used to aid in the design of systems incorporating atomic vapor in a cavity.  The present work constitutes part of a wider project in which the atom plus cavity system is needed to simultaneously enhance/suppress the density of states of the light field at certain frequencies: the modification of the cavity modes due to the contribution of the atoms to the cavity quickly become significant, or indeed surprising, as the number density of atoms is varied.  Furthermore, since the dispersion falls as \mbox{$1/\Delta$}, the cavity FSR is sensitive to the number and population distribution of the atomic ensemble.  This work highlights the challenges to reaching designed resonance conditions in optically thick optical cavities.  For example, in our case, changing the temperature by \mbox{$\sim1^\circ\mathrm{C}$} can shift the cavity response significantly, through the variation of the number density, pumping efficiency and thermal contributions to the width and shape of the response.

In this operational regime (elevated temperatures, reasonable buffer gas pressures) careful consideration of the broadening mechanisms is needed; in fact, having access to both the dispersion and absorption over a wide frequency range is also a useful diagnostic tool in characterising the key figures of merit for light-matter interactions of an atomic like system in a cavity, through a single shot measurement.

\section*{Acknowledgements}
We thank Eilon Poem, Amir Feizpour, and Krzysztof Kaczmarek for insightful discussions. This work was supported by the UK Engineering and Physical Sciences Research Council (EPSRC; Grant No. EP/J000051/1 and Programme Grant No. EP/K034480/1), the EPSRC Hub for Networked Quantum Information Technologies (NQIT), EU IP SIQS (Grant No. 600645), the Royal Society (J.N.), and EU Marie-Curie Fellowship Grant No. PIIF-GA-2013-629229 (D.J.S.). I.A.W. acknowledges an ERC Advanced Grant (MOQUACINO). C.Q. is supported by the China Scholarship Council (CSC No.201406140039) and J.H.D.M. is supported by the EPSRC Centre for Doctoral Training in Controlled Quantum Dynamics (under Grant No. EP/G037043/1).


\newpage
\begin{thebibliography}{29}%
\makeatletter
\providecommand \@ifxundefined [1]{%
 \@ifx{#1\undefined}
}%
\providecommand \@ifnum [1]{%
 \ifnum #1\expandafter \@firstoftwo
 \else \expandafter \@secondoftwo
 \fi
}%
\providecommand \@ifx [1]{%
 \ifx #1\expandafter \@firstoftwo
 \else \expandafter \@secondoftwo
 \fi
}%
\providecommand \natexlab [1]{#1}%
\providecommand \enquote  [1]{``#1''}%
\providecommand \bibnamefont  [1]{#1}%
\providecommand \bibfnamefont [1]{#1}%
\providecommand \citenamefont [1]{#1}%
\providecommand \href@noop [0]{\@secondoftwo}%
\providecommand \href [0]{\begingroup \@sanitize@url \@href}%
\providecommand \@href[1]{\@@startlink{#1}\@@href}%
\providecommand \@@href[1]{\endgroup#1\@@endlink}%
\providecommand \@sanitize@url [0]{\catcode `\\12\catcode `\$12\catcode
  `\&12\catcode `\#12\catcode `\^12\catcode `\_12\catcode `\%12\relax}%
\providecommand \@@startlink[1]{}%
\providecommand \@@endlink[0]{}%
\providecommand \url  [0]{\begingroup\@sanitize@url \@url }%
\providecommand \@url [1]{\endgroup\@href {#1}{\urlprefix }}%
\providecommand \urlprefix  [0]{URL }%
\providecommand \Eprint [0]{\href }%
\providecommand \doibase [0]{http://dx.doi.org/}%
\providecommand \selectlanguage [0]{\@gobble}%
\providecommand \bibinfo  [0]{\@secondoftwo}%
\providecommand \bibfield  [0]{\@secondoftwo}%
\providecommand \translation [1]{[#1]}%
\providecommand \BibitemOpen [0]{}%
\providecommand \bibitemStop [0]{}%
\providecommand \bibitemNoStop [0]{.\EOS\space}%
\providecommand \EOS [0]{\spacefactor3000\relax}%
\providecommand \BibitemShut  [1]{\csname bibitem#1\endcsname}%
\let\auto@bib@innerbib\@empty
\bibitem [{\citenamefont {Ladd}\ \emph {et~al.}(2010)\citenamefont {Ladd},
  \citenamefont {Jelezko}, \citenamefont {Laflamme}, \citenamefont {Nakamura},
  \citenamefont {Monroe},\ and\ \citenamefont {O’Brien}}]{Ladd2010}%
  \BibitemOpen
  \bibfield  {author} {\bibinfo {author} {\bibfnamefont {T.~D.}\ \bibnamefont
  {Ladd}}, \bibinfo {author} {\bibfnamefont {F.}~\bibnamefont {Jelezko}},
  \bibinfo {author} {\bibfnamefont {R.}~\bibnamefont {Laflamme}}, \bibinfo
  {author} {\bibfnamefont {Y.}~\bibnamefont {Nakamura}}, \bibinfo {author}
  {\bibfnamefont {C.}~\bibnamefont {Monroe}}, \ and\ \bibinfo {author}
  {\bibfnamefont {J.~L.}\ \bibnamefont {O’Brien}},\ }\href {\doibase
  10.1038/nature08812} {\bibfield  {journal} {\bibinfo  {journal} {Nature}\
  }\textbf {\bibinfo {volume} {464}},\ \bibinfo {pages} {45} (\bibinfo {year}
  {2010})}\BibitemShut {NoStop}%
\bibitem [{\citenamefont {Boozer}\ \emph {et~al.}(2007)\citenamefont {Boozer},
  \citenamefont {Boca}, \citenamefont {Miller}, \citenamefont {Northup},\ and\
  \citenamefont {Kimble}}]{Boozer2007}%
  \BibitemOpen
  \bibfield  {author} {\bibinfo {author} {\bibfnamefont {A.~D.}\ \bibnamefont
  {Boozer}}, \bibinfo {author} {\bibfnamefont {A.}~\bibnamefont {Boca}},
  \bibinfo {author} {\bibfnamefont {R.}~\bibnamefont {Miller}}, \bibinfo
  {author} {\bibfnamefont {T.~E.}\ \bibnamefont {Northup}}, \ and\ \bibinfo
  {author} {\bibfnamefont {H.~J.}\ \bibnamefont {Kimble}},\ }\href {\doibase
  10.1103/PhysRevLett.98.193601} {\bibfield  {journal} {\bibinfo  {journal}
  {Physical Review Letters}\ }\textbf {\bibinfo {volume} {98}},\ \bibinfo
  {pages} {193601} (\bibinfo {year} {2007})}\BibitemShut {NoStop}%
\bibitem [{\citenamefont {Saglamyurek}\ \emph {et~al.}(2011)\citenamefont
  {Saglamyurek}, \citenamefont {Sinclair}, \citenamefont {Jin}, \citenamefont
  {Slater}, \citenamefont {Oblak}, \citenamefont {Tittel}, \citenamefont
  {Bussi\`{e}res}, \citenamefont {George}, \citenamefont {Ricken},\ and\
  \citenamefont {Sohler}}]{Saglamyurek2011}%
  \BibitemOpen
  \bibfield  {author} {\bibinfo {author} {\bibfnamefont {E.}~\bibnamefont
  {Saglamyurek}}, \bibinfo {author} {\bibfnamefont {N.}~\bibnamefont
  {Sinclair}}, \bibinfo {author} {\bibfnamefont {J.}~\bibnamefont {Jin}},
  \bibinfo {author} {\bibfnamefont {J.~A.}\ \bibnamefont {Slater}}, \bibinfo
  {author} {\bibfnamefont {D.}~\bibnamefont {Oblak}}, \bibinfo {author}
  {\bibfnamefont {W.}~\bibnamefont {Tittel}}, \bibinfo {author} {\bibfnamefont
  {F.}~\bibnamefont {Bussi\`{e}res}}, \bibinfo {author} {\bibfnamefont
  {M.}~\bibnamefont {George}}, \bibinfo {author} {\bibfnamefont
  {R.}~\bibnamefont {Ricken}}, \ and\ \bibinfo {author} {\bibfnamefont
  {W.}~\bibnamefont {Sohler}},\ }\href {\doibase 10.1038/nature09719}
  {\bibfield  {journal} {\bibinfo  {journal} {Nature Letters}\ }\textbf
  {\bibinfo {volume} {469}},\ \bibinfo {pages} {512} (\bibinfo {year}
  {2011})},\ \Eprint {http://arxiv.org/abs/1009.0490} {arXiv:1009.0490}
  \BibitemShut {NoStop}%
\bibitem [{\citenamefont {Peyronel}\ \emph {et~al.}(2012)\citenamefont
  {Peyronel}, \citenamefont {Firstenberg}, \citenamefont {Liang}, \citenamefont
  {Hofferberth}, \citenamefont {Gorshkov}, \citenamefont {Pohl}, \citenamefont
  {Lukin},\ and\ \citenamefont {Vuleti\'{c}}}]{Peyronel2012}%
  \BibitemOpen
  \bibfield  {author} {\bibinfo {author} {\bibfnamefont {T.}~\bibnamefont
  {Peyronel}}, \bibinfo {author} {\bibfnamefont {O.}~\bibnamefont
  {Firstenberg}}, \bibinfo {author} {\bibfnamefont {Q.-Y.}\ \bibnamefont
  {Liang}}, \bibinfo {author} {\bibfnamefont {S.}~\bibnamefont {Hofferberth}},
  \bibinfo {author} {\bibfnamefont {A.~V.}\ \bibnamefont {Gorshkov}}, \bibinfo
  {author} {\bibfnamefont {T.}~\bibnamefont {Pohl}}, \bibinfo {author}
  {\bibfnamefont {M.~D.}\ \bibnamefont {Lukin}}, \ and\ \bibinfo {author}
  {\bibfnamefont {V.}~\bibnamefont {Vuleti\'{c}}},\ }\href {\doibase
  10.1038/nature11361} {\bibfield  {journal} {\bibinfo  {journal} {Nature}\
  }\textbf {\bibinfo {volume} {488}},\ \bibinfo {pages} {57} (\bibinfo {year}
  {2012})}\BibitemShut {NoStop}%
\bibitem [{\citenamefont {Hosseini}\ \emph {et~al.}(2011)\citenamefont
  {Hosseini}, \citenamefont {Campbell}, \citenamefont {Sparkes}, \citenamefont
  {Lam},\ and\ \citenamefont {Buchler}}]{Hosseini2011_nphys}%
  \BibitemOpen
  \bibfield  {author} {\bibinfo {author} {\bibfnamefont {M.}~\bibnamefont
  {Hosseini}}, \bibinfo {author} {\bibfnamefont {G.}~\bibnamefont {Campbell}},
  \bibinfo {author} {\bibfnamefont {B.~M.}\ \bibnamefont {Sparkes}}, \bibinfo
  {author} {\bibfnamefont {P.~K.}\ \bibnamefont {Lam}}, \ and\ \bibinfo
  {author} {\bibfnamefont {B.~C.}\ \bibnamefont {Buchler}},\ }\href {\doibase
  10.1038/nphys2021} {\bibfield  {journal} {\bibinfo  {journal} {Nature
  Physics}\ }\textbf {\bibinfo {volume} {7}},\ \bibinfo {pages} {794} (\bibinfo
  {year} {2011})}\BibitemShut {NoStop}%
\bibitem [{\citenamefont {Solja\v{c}i\'{c}}\ \emph {et~al.}(2005)\citenamefont
  {Solja\v{c}i\'{c}}, \citenamefont {Lidorikis}, \citenamefont {Hau},\ and\
  \citenamefont {Joannopoulos}}]{Soljacic2005}%
  \BibitemOpen
  \bibfield  {author} {\bibinfo {author} {\bibfnamefont {M.}~\bibnamefont
  {Solja\v{c}i\'{c}}}, \bibinfo {author} {\bibfnamefont {E.}~\bibnamefont
  {Lidorikis}}, \bibinfo {author} {\bibfnamefont {L.~V.}\ \bibnamefont {Hau}},
  \ and\ \bibinfo {author} {\bibfnamefont {J.~D.}\ \bibnamefont
  {Joannopoulos}},\ }\href {\doibase 10.1103/PhysRevE.71.026602} {\bibfield
  {journal} {\bibinfo  {journal} {Physical Review E}\ }\textbf {\bibinfo
  {volume} {71}},\ \bibinfo {pages} {026602} (\bibinfo {year}
  {2005})}\BibitemShut {NoStop}%
\bibitem [{\citenamefont {Lukin}\ \emph {et~al.}(1998)\citenamefont {Lukin},
  \citenamefont {Fleischhauer}, \citenamefont {Scully},\ and\ \citenamefont
  {Velichansky}}]{Lukin1998}%
  \BibitemOpen
  \bibfield  {author} {\bibinfo {author} {\bibfnamefont {M.~D.}\ \bibnamefont
  {Lukin}}, \bibinfo {author} {\bibfnamefont {M.}~\bibnamefont {Fleischhauer}},
  \bibinfo {author} {\bibfnamefont {M.~O.}\ \bibnamefont {Scully}}, \ and\
  \bibinfo {author} {\bibfnamefont {V.~L.}\ \bibnamefont {Velichansky}},\
  }\href@noop {} {\bibfield  {journal} {\bibinfo  {journal} {Optics letters}\
  }\textbf {\bibinfo {volume} {23}},\ \bibinfo {pages} {295} (\bibinfo {year}
  {1998})}\BibitemShut {NoStop}%
\bibitem [{\citenamefont {Wang}\ \emph {et~al.}(2000)\citenamefont {Wang},
  \citenamefont {Goorskey}, \citenamefont {Burkett},\ and\ \citenamefont
  {Xiao}}]{Wang2000}%
  \BibitemOpen
  \bibfield  {author} {\bibinfo {author} {\bibfnamefont {H.}~\bibnamefont
  {Wang}}, \bibinfo {author} {\bibfnamefont {D.~J.}\ \bibnamefont {Goorskey}},
  \bibinfo {author} {\bibfnamefont {W.~H.}\ \bibnamefont {Burkett}}, \ and\
  \bibinfo {author} {\bibfnamefont {M.}~\bibnamefont {Xiao}},\ }\href {\doibase
  10.1364/OL.25.001732} {\bibfield  {journal} {\bibinfo  {journal} {Optics
  letters}\ }\textbf {\bibinfo {volume} {25}},\ \bibinfo {pages} {1732}
  (\bibinfo {year} {2000})}\BibitemShut {NoStop}%
\bibitem [{\citenamefont {M\"{u}ller}\ \emph {et~al.}(1997)\citenamefont
  {M\"{u}ller}, \citenamefont {M\"{u}ller}, \citenamefont {Wicht},
  \citenamefont {Rinkleff},\ and\ \citenamefont {Danzmann}}]{Muller1997}%
  \BibitemOpen
  \bibfield  {author} {\bibinfo {author} {\bibfnamefont {G.}~\bibnamefont
  {M\"{u}ller}}, \bibinfo {author} {\bibfnamefont {M.}~\bibnamefont
  {M\"{u}ller}}, \bibinfo {author} {\bibfnamefont {A.}~\bibnamefont {Wicht}},
  \bibinfo {author} {\bibfnamefont {R.-H.}\ \bibnamefont {Rinkleff}}, \ and\
  \bibinfo {author} {\bibfnamefont {K.}~\bibnamefont {Danzmann}},\ }\href
  {\doibase 10.1103/PhysRevA.56.2385} {\bibfield  {journal} {\bibinfo
  {journal} {Physical Review A}\ }\textbf {\bibinfo {volume} {56}},\ \bibinfo
  {pages} {2385} (\bibinfo {year} {1997})}\BibitemShut {NoStop}%
\bibitem [{\citenamefont {Sabooni}\ \emph {et~al.}(2013)\citenamefont
  {Sabooni}, \citenamefont {Li}, \citenamefont {Rippe}, \citenamefont {Mohan},\
  and\ \citenamefont {Kr\"{o}ll}}]{Sabooni2013}%
  \BibitemOpen
  \bibfield  {author} {\bibinfo {author} {\bibfnamefont {M.}~\bibnamefont
  {Sabooni}}, \bibinfo {author} {\bibfnamefont {Q.}~\bibnamefont {Li}},
  \bibinfo {author} {\bibfnamefont {L.}~\bibnamefont {Rippe}}, \bibinfo
  {author} {\bibfnamefont {R.~K.}\ \bibnamefont {Mohan}}, \ and\ \bibinfo
  {author} {\bibfnamefont {S.}~\bibnamefont {Kr\"{o}ll}},\ }\href {\doibase
  10.1103/PhysRevLett.111.183602} {\bibfield  {journal} {\bibinfo  {journal}
  {Physical Review Letters}\ }\textbf {\bibinfo {volume} {111}},\ \bibinfo
  {pages} {1} (\bibinfo {year} {2013})},\ \Eprint
  {http://arxiv.org/abs/arXiv:1311.2301v1} {arXiv:arXiv:1311.2301v1}
  \BibitemShut {NoStop}%
\bibitem [{\citenamefont {Siddons}\ \emph {et~al.}(2009)\citenamefont
  {Siddons}, \citenamefont {Adams},\ and\ \citenamefont
  {Hughes}}]{Siddons2009}%
  \BibitemOpen
  \bibfield  {author} {\bibinfo {author} {\bibfnamefont {P.}~\bibnamefont
  {Siddons}}, \bibinfo {author} {\bibfnamefont {C.~S.}\ \bibnamefont {Adams}},
  \ and\ \bibinfo {author} {\bibfnamefont {I.~G.}\ \bibnamefont {Hughes}},\
  }\href {\doibase 10.1088/0953-4075/42/17/175004} {\bibfield  {journal}
  {\bibinfo  {journal} {Journal of Physics B: Atomic, Molecular and Optical
  Physics}\ }\textbf {\bibinfo {volume} {42}},\ \bibinfo {pages} {175004}
  (\bibinfo {year} {2009})}\BibitemShut {NoStop}%
\bibitem [{\citenamefont {Schultz}\ \emph {et~al.}(2008)\citenamefont
  {Schultz}, \citenamefont {Ming}, \citenamefont {Noble},\ and\ \citenamefont
  {van Wijngaarden}}]{Schultz2008}%
  \BibitemOpen
  \bibfield  {author} {\bibinfo {author} {\bibfnamefont {B.~E.}\ \bibnamefont
  {Schultz}}, \bibinfo {author} {\bibfnamefont {H.}~\bibnamefont {Ming}},
  \bibinfo {author} {\bibfnamefont {G.~A.}\ \bibnamefont {Noble}}, \ and\
  \bibinfo {author} {\bibfnamefont {W.~A.}\ \bibnamefont {van Wijngaarden}},\
  }\href {\doibase 10.1140/epjd/e2008-00109-0} {\bibfield  {journal} {\bibinfo
  {journal} {The European Physical Journal D}\ }\textbf {\bibinfo {volume}
  {48}},\ \bibinfo {pages} {171} (\bibinfo {year} {2008})}\BibitemShut
  {NoStop}%
\bibitem [{\citenamefont {Saunders}\ \emph {et~al.}()\citenamefont {Saunders},
  \citenamefont {Munns}, \citenamefont {Champion}, \citenamefont {Qiu},
  \citenamefont {Kaczmarek}, \citenamefont {Poem}, \citenamefont {Ledingham},
  \citenamefont {Walmsley},\ and\ \citenamefont {Nunn}}]{DJS2015}%
  \BibitemOpen
  \bibfield  {author} {\bibinfo {author} {\bibfnamefont {D.~J.}\ \bibnamefont
  {Saunders}}, \bibinfo {author} {\bibfnamefont {J.~H.~D.}\ \bibnamefont
  {Munns}}, \bibinfo {author} {\bibfnamefont {T.~F.~M.}\ \bibnamefont
  {Champion}}, \bibinfo {author} {\bibfnamefont {C.}~\bibnamefont {Qiu}},
  \bibinfo {author} {\bibfnamefont {K.~T.}\ \bibnamefont {Kaczmarek}}, \bibinfo
  {author} {\bibfnamefont {E.}~\bibnamefont {Poem}}, \bibinfo {author}
  {\bibfnamefont {P.~M.}\ \bibnamefont {Ledingham}}, \bibinfo {author}
  {\bibfnamefont {I.~A.}\ \bibnamefont {Walmsley}}, \ and\ \bibinfo {author}
  {\bibfnamefont {J.}~\bibnamefont {Nunn}},\ }\href
  {http://arxiv.org/abs/1510.04625} {\ }\Eprint
  {http://arxiv.org/abs/1510.04625} {arXiv:1510.04625} \BibitemShut {NoStop}%
\bibitem [{\citenamefont {Keaveney}(2014)}]{Keaveney2014_thesis}%
  \BibitemOpen
  \bibfield  {author} {\bibinfo {author} {\bibfnamefont {J.}~\bibnamefont
  {Keaveney}},\ }\href {\doibase 10.1007/978-3-319-07100-8} {\emph {\bibinfo
  {title} {{Collective Atom – Light Interactions in Dense Atomic Vapours}}}}\
  (\bibinfo  {publisher} {Springer},\ \bibinfo {year} {2014})\BibitemShut
  {NoStop}%
\bibitem [{\citenamefont {Zentile}\ \emph {et~al.}(2015)\citenamefont
  {Zentile}, \citenamefont {Keaveney}, \citenamefont {Weller}, \citenamefont
  {Whiting}, \citenamefont {Adams},\ and\ \citenamefont
  {Hughes}}]{Zentile2015}%
  \BibitemOpen
  \bibfield  {author} {\bibinfo {author} {\bibfnamefont {M.~A.}\ \bibnamefont
  {Zentile}}, \bibinfo {author} {\bibfnamefont {J.}~\bibnamefont {Keaveney}},
  \bibinfo {author} {\bibfnamefont {L.}~\bibnamefont {Weller}}, \bibinfo
  {author} {\bibfnamefont {D.~J.}\ \bibnamefont {Whiting}}, \bibinfo {author}
  {\bibfnamefont {C.~S.}\ \bibnamefont {Adams}}, \ and\ \bibinfo {author}
  {\bibfnamefont {I.~G.}\ \bibnamefont {Hughes}},\ }\href {\doibase
  10.1016/j.cpc.2014.11.023} {\bibfield  {journal} {\bibinfo  {journal}
  {Computer Physics Communications}\ }\textbf {\bibinfo {volume} {189}},\
  \bibinfo {pages} {162} (\bibinfo {year} {2015})}\BibitemShut {NoStop}%
\bibitem [{\citenamefont {Siddons}\ \emph {et~al.}(2008)\citenamefont
  {Siddons}, \citenamefont {Adams}, \citenamefont {Ge},\ and\ \citenamefont
  {Hughes}}]{Siddons2008}%
  \BibitemOpen
  \bibfield  {author} {\bibinfo {author} {\bibfnamefont {P.}~\bibnamefont
  {Siddons}}, \bibinfo {author} {\bibfnamefont {C.~S.}\ \bibnamefont {Adams}},
  \bibinfo {author} {\bibfnamefont {C.}~\bibnamefont {Ge}}, \ and\ \bibinfo
  {author} {\bibfnamefont {I.~G.}\ \bibnamefont {Hughes}},\ }\href {\doibase
  10.1088/0953-4075/41/15/155004} {\bibfield  {journal} {\bibinfo  {journal}
  {Journal of Physics B: Atomic, Molecular and Optical Physics}\ }\textbf
  {\bibinfo {volume} {41}},\ \bibinfo {pages} {155004} (\bibinfo {year}
  {2008})}\BibitemShut {NoStop}%
\bibitem [{\citenamefont {Steck}(2003)}]{Steck2003}%
  \BibitemOpen
  \bibfield  {author} {\bibinfo {author} {\bibfnamefont {D.~A.}\ \bibnamefont
  {Steck}},\ }\href {http://steck.us/alkalidata} {\enquote {\bibinfo {title}
  {{Cesium D Line Data}},}\ } (\bibinfo {year} {2003})\BibitemShut {NoStop}%
\bibitem [{\citenamefont {Loudon}(2000)}]{Loudon2000_qtol}%
  \BibitemOpen
  \bibfield  {author} {\bibinfo {author} {\bibfnamefont {R.}~\bibnamefont
  {Loudon}},\ }\href@noop {} {\emph {\bibinfo {title} {{The Quantum Theory of
  Light}}}},\ \bibinfo {edition} {3rd}\ ed.\ (\bibinfo  {publisher} {OUP
  Oxford},\ \bibinfo {year} {2000})\BibitemShut {NoStop}%
\bibitem [{\citenamefont {Pitz}\ \emph {et~al.}(2010)\citenamefont {Pitz},
  \citenamefont {Fox},\ and\ \citenamefont {Perram}}]{Pitz2010}%
  \BibitemOpen
  \bibfield  {author} {\bibinfo {author} {\bibfnamefont {G.~A.}\ \bibnamefont
  {Pitz}}, \bibinfo {author} {\bibfnamefont {C.~D.}\ \bibnamefont {Fox}}, \
  and\ \bibinfo {author} {\bibfnamefont {G.~P.}\ \bibnamefont {Perram}},\
  }\href {\doibase 10.1103/PhysRevA.82.042502} {\bibfield  {journal} {\bibinfo
  {journal} {Physical Review A}\ }\textbf {\bibinfo {volume} {82}},\ \bibinfo
  {pages} {042502} (\bibinfo {year} {2010})}\BibitemShut {NoStop}%
\bibitem [{\citenamefont {Kozlova}\ \emph {et~al.}(2011)\citenamefont
  {Kozlova}, \citenamefont {Gu\'{e}randel},\ and\ \citenamefont
  {de~Clercq}}]{Kozlova2011}%
  \BibitemOpen
  \bibfield  {author} {\bibinfo {author} {\bibfnamefont {O.}~\bibnamefont
  {Kozlova}}, \bibinfo {author} {\bibfnamefont {S.}~\bibnamefont
  {Gu\'{e}randel}}, \ and\ \bibinfo {author} {\bibfnamefont {E.}~\bibnamefont
  {de~Clercq}},\ }\href {\doibase 10.1103/PhysRevA.83.062714} {\bibfield
  {journal} {\bibinfo  {journal} {Physical Review A}\ }\textbf {\bibinfo
  {volume} {83}},\ \bibinfo {pages} {062714} (\bibinfo {year}
  {2011})}\BibitemShut {NoStop}%
\bibitem [{\citenamefont {Thompson}\ \emph {et~al.}(1987)\citenamefont
  {Thompson}, \citenamefont {Cox},\ and\ \citenamefont
  {Hastings}}]{Thompson1986}%
  \BibitemOpen
  \bibfield  {author} {\bibinfo {author} {\bibfnamefont {P.}~\bibnamefont
  {Thompson}}, \bibinfo {author} {\bibfnamefont {D.~E.}\ \bibnamefont {Cox}}, \
  and\ \bibinfo {author} {\bibfnamefont {J.~B.}\ \bibnamefont {Hastings}},\
  }\href@noop {} {\bibfield  {journal} {\bibinfo  {journal} {J. Appl. Cryst.}\
  }\textbf {\bibinfo {volume} {20}},\ \bibinfo {pages} {79} (\bibinfo {year}
  {1987})}\BibitemShut {NoStop}%
\bibitem [{\citenamefont {Wells}(1999)}]{Wells1999}%
  \BibitemOpen
  \bibfield  {author} {\bibinfo {author} {\bibfnamefont {R.~J.}\ \bibnamefont
  {Wells}},\ }\href {\doibase 10.1016/S0022-4073(97)00231-8} {\bibfield
  {journal} {\bibinfo  {journal} {Journal of Quantitative Spectroscopy and
  Radiative Transfer}\ }\textbf {\bibinfo {volume} {62}},\ \bibinfo {pages}
  {29} (\bibinfo {year} {1999})}\BibitemShut {NoStop}%
\bibitem [{\citenamefont {Johnson}(2012)}]{SGJohnson2012}%
  \BibitemOpen
  \bibfield  {author} {\bibinfo {author} {\bibfnamefont {S.~G.}\ \bibnamefont
  {Johnson}},\ }\href
  {http://ab-initio.mit.edu/wiki/index.php/Faddeeva\_Package} {\enquote
  {\bibinfo {title} {{Faddeeva Package}},}\ } (\bibinfo {year}
  {2012})\BibitemShut {NoStop}%
\bibitem [{\citenamefont {Wicht}\ \emph {et~al.}(2000)\citenamefont {Wicht},
  \citenamefont {Muller}, \citenamefont {Rinkleff}, \citenamefont {Rocco},\
  and\ \citenamefont {Danzmann}}]{Wicht2000}%
  \BibitemOpen
  \bibfield  {author} {\bibinfo {author} {\bibfnamefont {A.}~\bibnamefont
  {Wicht}}, \bibinfo {author} {\bibfnamefont {M.}~\bibnamefont {Muller}},
  \bibinfo {author} {\bibfnamefont {R.-H.}\ \bibnamefont {Rinkleff}}, \bibinfo
  {author} {\bibfnamefont {A.}~\bibnamefont {Rocco}}, \ and\ \bibinfo {author}
  {\bibfnamefont {K.}~\bibnamefont {Danzmann}},\ }\href@noop {} {\bibfield
  {journal} {\bibinfo  {journal} {Optics Communications}\ }\textbf {\bibinfo
  {volume} {179}},\ \bibinfo {pages} {107} (\bibinfo {year}
  {2000})}\BibitemShut {NoStop}%
\bibitem [{\citenamefont {Reim}(2011)}]{Reim2011_thesis}%
  \BibitemOpen
  \bibfield  {author} {\bibinfo {author} {\bibfnamefont {K.~F.}\ \bibnamefont
  {Reim}},\ }\emph {\bibinfo {title} {{Broadband optical quantum memory}}},\
  \href@noop {} {Ph.D. thesis},\ \bibinfo  {school} {University of Oxford}
  (\bibinfo {year} {2011})\BibitemShut {NoStop}%
\bibitem [{\citenamefont {Sprague}\ \emph {et~al.}(2014)\citenamefont
  {Sprague}, \citenamefont {Michelberger}, \citenamefont {Champion},
  \citenamefont {England}, \citenamefont {Nunn}, \citenamefont {Jin},
  \citenamefont {Kolthammer}, \citenamefont {Abdolvand}, \citenamefont
  {Russell},\ and\ \citenamefont {Walmsley}}]{Sprague2014}%
  \BibitemOpen
  \bibfield  {author} {\bibinfo {author} {\bibfnamefont {M.~R.}\ \bibnamefont
  {Sprague}}, \bibinfo {author} {\bibfnamefont {P.~S.}\ \bibnamefont
  {Michelberger}}, \bibinfo {author} {\bibfnamefont {T.~F.~M.}\ \bibnamefont
  {Champion}}, \bibinfo {author} {\bibfnamefont {D.~G.}\ \bibnamefont
  {England}}, \bibinfo {author} {\bibfnamefont {J.}~\bibnamefont {Nunn}},
  \bibinfo {author} {\bibfnamefont {X.-M.}\ \bibnamefont {Jin}}, \bibinfo
  {author} {\bibfnamefont {W.~S.}\ \bibnamefont {Kolthammer}}, \bibinfo
  {author} {\bibfnamefont {A.}~\bibnamefont {Abdolvand}}, \bibinfo {author}
  {\bibfnamefont {P.~S.~J.}\ \bibnamefont {Russell}}, \ and\ \bibinfo {author}
  {\bibfnamefont {I.~A.}\ \bibnamefont {Walmsley}},\ }\href {\doibase
  10.1038/nphoton.2014.45} {\bibfield  {journal} {\bibinfo  {journal} {Nature
  Photonics}\ }\textbf {\bibinfo {volume} {8}},\ \bibinfo {pages} {287}
  (\bibinfo {year} {2014})}\BibitemShut {NoStop}%
\bibitem [{\citenamefont {Weller}\ \emph {et~al.}(2011)\citenamefont {Weller},
  \citenamefont {Bettles}, \citenamefont {Siddons}, \citenamefont {Adams},\
  and\ \citenamefont {Hughes}}]{Weller2011}%
  \BibitemOpen
  \bibfield  {author} {\bibinfo {author} {\bibfnamefont {L.}~\bibnamefont
  {Weller}}, \bibinfo {author} {\bibfnamefont {R.~J.}\ \bibnamefont {Bettles}},
  \bibinfo {author} {\bibfnamefont {P.}~\bibnamefont {Siddons}}, \bibinfo
  {author} {\bibfnamefont {C.~S.}\ \bibnamefont {Adams}}, \ and\ \bibinfo
  {author} {\bibfnamefont {I.~G.}\ \bibnamefont {Hughes}},\ }\href {\doibase
  10.1088/0953-4075/44/19/195006} {\bibfield  {journal} {\bibinfo  {journal}
  {J. Phys. B.}\ }\textbf {\bibinfo {volume} {195006}} (\bibinfo {year}
  {2011}),\ 10.1088/0953-4075/44/19/195006},\ \Eprint
  {http://arxiv.org/abs/1107.3092} {arXiv:1107.3092} \BibitemShut {NoStop}%
\bibitem [{\citenamefont {Hansch}\ and\ \citenamefont
  {Couillaud}(1980)}]{Hansch1980}%
  \BibitemOpen
  \bibfield  {author} {\bibinfo {author} {\bibfnamefont {T.~W.}\ \bibnamefont
  {Hansch}}\ and\ \bibinfo {author} {\bibfnamefont {B.}~\bibnamefont
  {Couillaud}},\ }\href@noop {} {\bibfield  {journal} {\bibinfo  {journal}
  {Optical Communications}\ }\textbf {\bibinfo {volume} {35}},\ \bibinfo
  {pages} {441} (\bibinfo {year} {1980})}\BibitemShut {NoStop}%
\bibitem [{\citenamefont {Rosenberry}\ \emph {et~al.}(2007)\citenamefont
  {Rosenberry}, \citenamefont {Reyes}, \citenamefont {Tupa},\ and\
  \citenamefont {Gay}}]{Rosenberry2007}%
  \BibitemOpen
  \bibfield  {author} {\bibinfo {author} {\bibfnamefont {M.~A.}\ \bibnamefont
  {Rosenberry}}, \bibinfo {author} {\bibfnamefont {J.~P.}\ \bibnamefont
  {Reyes}}, \bibinfo {author} {\bibfnamefont {D.}~\bibnamefont {Tupa}}, \ and\
  \bibinfo {author} {\bibfnamefont {T.~J.}\ \bibnamefont {Gay}},\ }\href
  {\doibase 10.1103/PhysRevA.75.023401} {\bibfield  {journal} {\bibinfo
  {journal} {Physical Review A}\ }\textbf {\bibinfo {volume} {75}},\ \bibinfo
  {pages} {1} (\bibinfo {year} {2007})}\BibitemShut {NoStop}%
\end{thebibliography}
\end{document}